\documentclass[conference]{IEEEtran}
\usepackage[OT1]{fontenc} 
\usepackage{cite}
\usepackage{amsmath,amssymb,amsfonts}
\usepackage{graphicx}
\usepackage{textcomp}
\usepackage{xcolor}
\usepackage{url}
\usepackage{txfonts}
\usepackage{bm}
\usepackage{algorithm}
\usepackage{algpseudocode}
\usepackage{booktabs}
\usepackage{tabu}
\usepackage{float}

\def\BibTeX{{\rm B\kern-.05em{\sc i\kern-.025em b}\kern-.08em
    T\kern-.1667em\lower.7ex\hbox{E}\kern-.125emX}}
\begin{document}

\title{Domain-Embeddings Based DGA Detection with Incremental Training Method\\
}

\author{
\IEEEauthorblockN{1\textsuperscript{st} Xin Fang, 2\textsuperscript{nd} Xiaoqing Sun,3\textsuperscript{rd} Jiahai Yang}
\IEEEauthorblockA{\textit{Institute for Network Sciences and Cyberspace} \\
\textit{Tsinghua University}, Beijing, China \\
Beijing National Research Center for Information \\
Science and Technology\\
\{fx18, sxq16\}@mails.tsinghua.edu.cn, yang@cernet.edu.cn}

\and
\IEEEauthorblockN{4\textsuperscript{th} Xinran Liu}
\IEEEauthorblockA{\textit{National Computer Network Emergency Response Technical} \\
\textit{Team/Coordination Center}, Beijing, China\\
lxr@cert.org.cn}
}
\IEEEoverridecommandlockouts
\IEEEpubid{\makebox[\columnwidth]{978-1-7281-8086-1/20/\$31.00~\copyright2020 IEEE \hfill} \hspace{\columnsep}\makebox[\columnwidth]{ }}

\maketitle

\IEEEpubidadjcol
\begin{abstract}

DGA-based botnet, which uses Domain Generation Algorithms (DGAs) to evade supervision,  has become a part of the most destructive threats to network security. Over the past decades, a wealth of defense mechanisms focusing on domain features have emerged to address the problem. Nonetheless, DGA detection remains a daunting and challenging task due to the big data nature of Internet traffic and the potential fact that the linguistic features extracted only from the domain names are insufficient and the enemies could easily forge them to disturb detection.
In this paper, we propose a novel DGA detection system which employs an incremental word-embeddings method to capture the interactions between end hosts and domains, characterize time-series patterns of DNS queries for each IP address and therefore explore temporal similarities between domains. We carefully modify the Word2Vec algorithm and leverage it to automatically learn dynamic and discriminative feature representations for over 1.9 million domains, and develop an simple classifier for distinguishing malicious domains from the benign.
Given the ability to identify temporal patterns of domains and update models incrementally, the proposed scheme makes the progress towards adapting to the changing and evolving strategies of DGA domains.
Our system is evaluated and compared with the state-of-art system FANCI and two deep-learning methods CNN and LSTM, with data from a large university's network named TUNET. The results suggest that our system outperforms the strong competitors by a large margin on multiple metrics and meanwhile achieves a remarkable speed-up on model updating.

\end{abstract}


\section{Introduction}

DGAs are commonly used by botnets to bypass security mechanisms where some static methods like blacklists are employed. They can generate a vast amount of pseudo-random domain names, while the attacker would only select a small subset for registration to establish command and control (C\&C) connections \cite{FromThrowAwayTraffictoBots}  \cite{AComprehensiveMeasurementStudyofDomainGeneratingMalware}.This results in an asymmetric situation where attackers can use any one of generated domains to control bots, but defenders must monitor all of them.
A wide spectrum of methods for DGA detection have been proposed in recent years, but most of them rely upon the linguistic features developed and could not work well when the botmaster decides to change domain-generating strategy. This explains why some character-based detectors which work well for traditional DGAs  perform poorly when confronted with those plausibly clean-looking domain names based on wordlists (also called dictionaries) \cite{DictionaryExtraction}.

In such scenario, we are concerned with developing an algorithm that is resilient to feature change and able to function well for not only character-based or wordlist-based DGAs , but also for any kind of completely new algorithms that are never seen before.
We hold the intuition that bots tend to exhibit similar behavior patterns no matter what kind of DGA algorithms are implemented. These time-relevant patterns provide more robust and stable features which improve the flexibility against the changing and evolving attacking strategies.
For example, the bots controlled by the same entity communicate with the same C\&C server, and the botnet members cause a large amount of regular traffic when launching an attack. 
In addition, our system must be adaptive to the big data nature of Internet traffic and the explosive growth of malicious domains, so an well-designed incremental training strategy is indispensable to reduce the model iteration cost.

In this paper, we propose a novel DGA detection system aiming for a wide range of DGA families and the never-ending growth of the Internet DNS traffic. The critical nature of our ideas is to characterize time-series patterns of DNS queries for each IP address, explore temporal similarities between domains and apply incremental training strategy to speed up model updating.

To sum up, the contributions of this work is threefold:

\textbf{1)} In order to improve the flexibility against the changing and evolving attacking strategies, we focus on the underlying relevance among the domains and utilize the latent patterns of DNS query sequences to detect DGAs. We also apply word2vec algorithm for a mapping from DGA detection to vector arithmetic.

\textbf{2)} To cope with the never-ending growth of the Internet DNS traffic, we utilize an incremental training strategy for word2vec al- gorithm, which helps to speed up the model training process when additional training data is provided.

\textbf{3)} We built a practical system based on the proposed algorithm, and achieved excellent results in several empirical experiments and real-world deployments.

The rest of this paper is organized as follows. In Section 2, we introduce some background knowledge and systematically outline related works. In Section 3, we propose our DGA detection system based on the incremental word2vec algorithm with details. Then we provide our experimental methodology and results in Section 4. Finally, we summarize our primary jobs and discuss future work in Section 5.

\section{Related Work}

\subsection{DGA and DGA Detection}
In order to detect DGA domains, Yadav et al. \cite{Yadav2010Detecting} proposed a technique based on the significant difference between traditional DGA domains and human generated domains in terms of the distribution of alphanumeric characters. In addition, Antonakakis et al. \cite{FromThrowAwayTraffictoBots}, Sch{\"u}ppen et al. \cite{schuppen2018fanci} and Wang et al. \cite{SVMdetector} proposed machine-learning based DGA detectors using human-engineered lexical features of DGA domain names, while Tong et al. \cite{d3n}, Lison et al. \cite{RecurrentNeuralModels} and Tran et al. \cite{LSTMDGA} came up with some methods using deep learning algorithms such as CNN, LSTM, and BiLSTM. However, attackers have designed a more resilient class of mAGDs produced by randomly selecting and concatenating words from a dictionary in order to imitate legitimate domain names created by a human. This new kind of DGA is much harder to detect. In fact, many state-of-the-art DGA detectors which function well for traditional DGAs, perform poorly when faced with wordlist-based ones.

Confronted with such a challenging situation, defenders have presented several countermeasures. Pereira et al. \cite{DictionaryExtraction} firstly proposed a method for combating wordlist-based DGAs in 2018. They built a new structure named WordGraph based on the segmentation of domain names, and then employ it to further discover DGA dictionaries. Another existing approach raised by J.Koh et al. \cite{InlineDetection} extracted in-depth semantic features from unrelated corpus, and used \emph{transfer learning} theory to learn the semantic signatures of the wordlist-based DGA families. These approaches perform well, but only focus on wordlist-based DGAs nevertheless.

\subsection{Word Embeddings Algorithm}
The basic idea of \emph{word embeddings} was initially proposed by Hinton in 1986 \cite{hinton1986learning}, which was called \emph{distributed representation} at that time. This method is mainly used in the area of Natural Language Processing (NLP), while we can still utilize it in Domain Name System (DNS) analyzing field by analogizing DNS query sequences to natural language sentences, which follows the core idea of W. Lopez et al. \cite{DNSwordembedding} that considers DNS queries from a particular source IP address during a specific time interval as \emph{words} in a single \emph{document}. Nowadays the most ubiquitous word embeddings method is \emph{Word2Vec} \cite{Mikolov2013EfficientEO}, and in this paper we use \emph{Skip-Gram model with Negative Sampling (SGNS)} \cite{SGNS}, an advanced variant of \emph{Word2Vec}, as basic algorithm for its popularity. Several previous works tried to apply word2vec algorithm to DNS- related field ( \cite{Dns2Vec}, \cite{Domain2Vec}), but their target is traffic classification instead of DGA detection.

\begin{figure}[htbp]
\centerline{\includegraphics[width=3.5in]{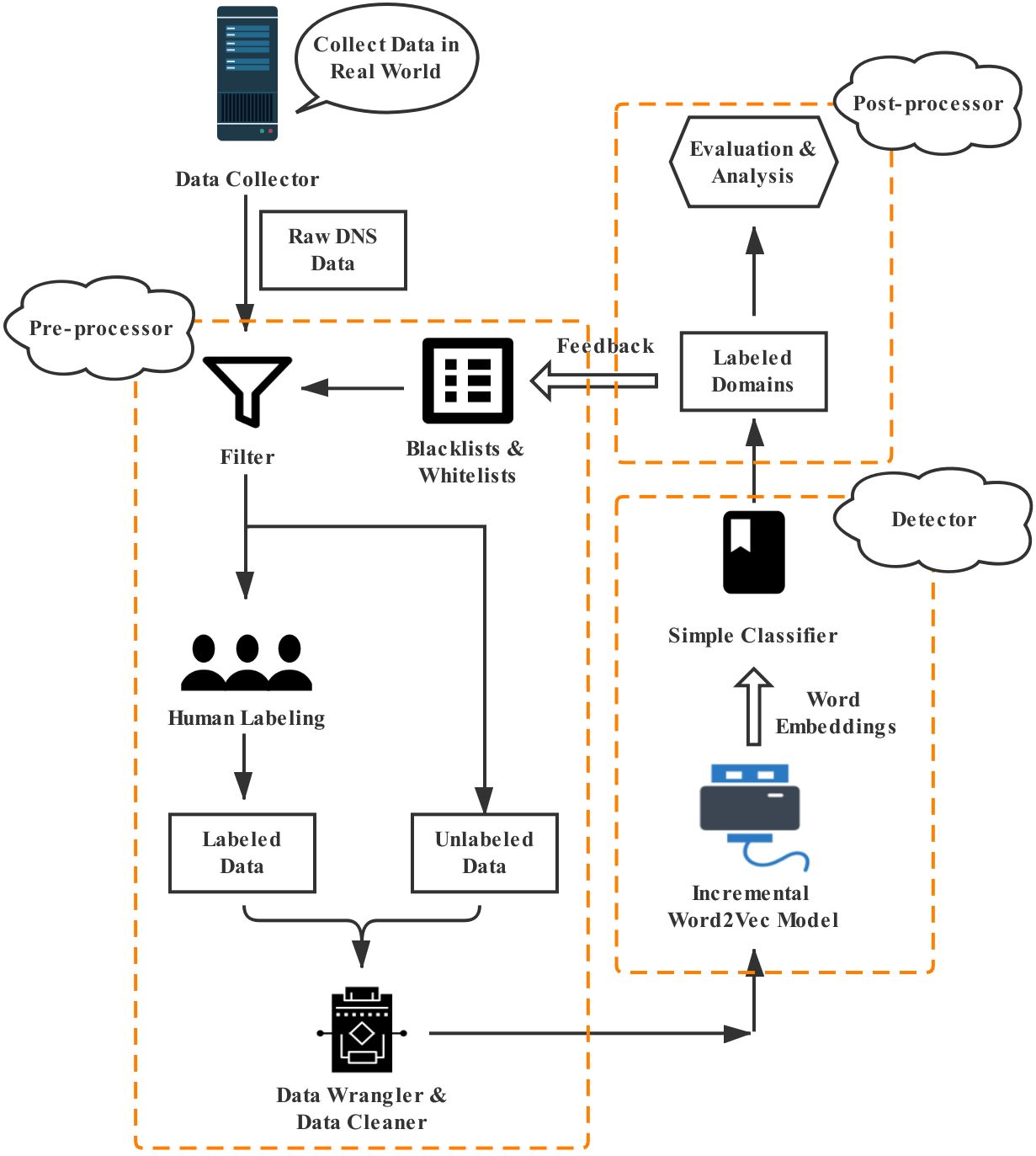}}
\caption{System architecture.}
\label{fig}
\end{figure}

Although we illustrate that SGNS can accurately classify mAGDs through experiments, it turns out that existing neural word embeddings methods, including SGNS, are multi-pass algorithms and thus cannot perform incremental model update, which means that they have to re-train the model on the old and new training data from scratch when additional training data is provided \cite{IncrementalSGNS}. To this end, some researchers have focused on exploring incremental training strategies of word embeddings methods ( \cite{IncrementalSGNS}, \cite{spacesaving}). Similar to the conventional word2vec algorithm, there was also little or no attention paid to the incremental word2vec method in the literature when it comes to DGA detection as far as we know.

\section{System Architecture}
In this section, we describe our DGA detection system architecture and training mechanism. As is shown in Fig. 1, our system has several components: Pre-processor, Detector, and Post-processor.

\subsection{Pre-processor}
Before the preprocessing phase, we deploy our data collectors in several core DNS servers in TUNET to collect raw DNS logs. Thereafter, we apply black/white lists from both public sources (e.g., \emph{malwaredomainlist.com}) and private sources to raw DNS corpus, which can be considered as a pre-labeling process. To further calibrate the labeling results, we sample part of raw data for manual labeling. As for the unlabeled leftovers, they will be used for training word embeddings since word2vec algorithm is unsupervised.

After labeling process, we feed all of the data to a data wrangling and cleaning module, which functions as follows: First of all, the module traverses through the entire dataset and remove all queries containing invalid IP addresses, query type or query name. Second, since many of the DNS queries are nonexistent domain names, rarely duplicated, and in many cases composed of a large number of changing prefixes and a few unchanging suffixes, we decide to merge the similar domains by common suffixes. Besides, to eliminate the impact of ccTLD (country code Top-Level Domain\cite{ccTLD}), we remove all ccTLDs from the tail of domain names containing them. Last but not least, we select the appropriate time window size and reorganize the data structure. More specifically, we determine a window size such as 10 minutes, which remains as a hyper-parameter to be decided later and partition the dataset accordingly. Query records in each window are organized in the format of [$timestamp$, $IP$, $domain_{1}$, $domain_{2}$, $\cdots$]. Finally, all queries from a specific IP address during the pre-defined time window are grouped and hence constitute a \emph{Document} with each domain name is a \emph{Word}.

\subsection{Detector}
Our detector consists of two parts: the incremental word2vec model and the subsequent simple classifier.

\subsubsection{Incremental Word2Vec Model}
Based on previous research results (\cite{ IncrementalSGNS}, \cite{spacesaving}), we apply the incremental training method of  SGNS to the domain-embeddings generation model in this paper. Given one document output from Pre-processor, we assume that the words (domains) inside constitute a sequence: $w_{1}$, $w_{2}$, $w_{3}$, $\cdots$ , $w_{n}$. Then the classical SGNS model attempts to minimize the following objective function to learn domain embeddings: 

\begin{small}
\begin{equation}
\mathcal{L}_{SGNS} = -\frac{1}{n}\sum_{i=1}^{n}\sum_{|j|<c, j\neq0}log \sigma(\mathbf{t}_{w_{i}}\cdot\mathbf{c}_{w_{i+j}}) + k{\cal{E}}_{v \sim q(v)}[log\sigma(-\mathbf{t}_{w_{i}}\cdot\mathbf{c}_{v}]
\end{equation}
\end{small}

where $\mathbf{t}_{w_{i}}$ is the target word $w_{i}$'s embedding and $\mathbf{c}_{w_{i+j}}$ is the context word $w_{i+j}$'s embedding within a window of size $c$, $\sigma(x)$ is the sigmoid function, $k$ is a pre-fixed integer, $v$ is the negative sample drawn from $q(v)$, and $q(v)$, which is referred to as \emph{negative sampling distribution}\cite{spacesaving}. While Equation(1) can be optimized by Stochastic Gradient Descent (SGD) using AdaGrad\cite{duchi2011adaptive} in an online fashion, traditional multi-pass SGNS training still needs to scan through the entire dataset at first to pre-compute the negative sampling distribution $q(v)$ , which makes it difficult to perform efficient incremental model update when additional training data is provided every single time, especially when the amount of the new data is smaller compared to the old one.

For the reason that new domains keep showing up continuously in real world, we need to present an incremental extension of SGNS. We adopt the methodology inherited from previous works (\cite{IncrementalSGNS}, \cite{spacesaving}), which goes through the training data solely in a single-pass to update word embeddings incrementally.  Algorithm 1 presents this incremental SGNS algorithm.

\begin{algorithm}[htbp]
  \caption{Incremental SGNS}
  \begin{algorithmic}[1]
	\For{each new batch $\cal{D}$ of training data}	
	  \State $f(d) \leftarrow 0$ for all $d \in \cal{D}$
	  \State $n \leftarrow length(\cal{D})$
	  \For {$i \leftarrow 1$, $\cdots$, $n$}
	  	\State $f(d_{i}) \leftarrow f(d_{i}) + 1$
	  	\State $q(d) \leftarrow \frac{f(d)^\alpha}{\Sigma_{d' \in \cal{D}}f(d')^\alpha}$ for all $d \in \cal{D}$
	  	\For {$j \leftarrow -c, \cdots, -1, 1, \cdots, c$}
	  		\State draw $k$ negative samples from $q(d)$
	  		\State use adaptive SGD to update $\mathbf{t}_{w_{i}}$, $\mathbf{c}_{w_{i+j}}$, and $c_{v_{1}}, \cdots, c_{v_{k}}$
	  	\EndFor
	  \EndFor
	\EndFor
  \end{algorithmic}
\end{algorithm}

\begin{algorithm}[htbp]
  \caption{Draw Negative Samples}
  \begin{algorithmic}[1]
  	\State set array $r$ with length $K$ to empty
  	\State $n \leftarrow length(\cal{W})$
  	\State $cnt \leftarrow 0$
  	\For {$i \leftarrow 1$, $\cdots$, $n$}
  	  \State $cnt \leftarrow cnt + 1$
  	  \If {$i \leq K$}
  	  	\State $r_{i} \leftarrow w_{i}$
  	  \Else
  	  	\State draw an interger $k$ uniformly from ${1,2,\cdots,n}$
  	  	\If {$k \leq K$}
  	  	  \State $r_{k} \leftarrow w_{i}$
  	  	\EndIf
  	  \EndIf
  	\EndFor
  \end{algorithmic}
\end{algorithm}

In the implementation of incremental SGNS, how to efficiently produce negative samples is an important issue, since the efficiency of sampling greatly affects the overall training speed. To seek solution to this problem, here we utilize the Reservoir Sampling \cite{vitter1985random} algorithm, which helps to generate one single negative sample in only \emph{O(1)} time (See Algorithm 2).

\subsubsection{Logistic Regression Classifier}
Without loss of generality, we use the Logistic Regression classifier as the tail classifier. It receives all the word-embeddings popped out from word2vec model and the corresponding ground-truth label, which specifies whether the domain is malicious or not. It is noteworthy that logistic regression already has the potential for incremental training because it can update the parameters using SGD every time there is new training data provided. In the testing/evaluation/deploy phase, the classifier can directly calculate the input domain-embeddings' labels without extra operations.

\subsubsection{Workflow Description}
Before Detector, we already di- vide datasets into labeled part (relatively small) and unlabelled part (relatively big) due to expensive manual labeling cost. Fortunately, we do not need ground truth when training unsupervised word2vec models. Hence, our overall training strategy is to use all received valid data for training word2vec model, while feed only labeled ones to the classifier to make the best use of collected data. Actually, based on such a greedy strategy, we have guaranteed the quality and generalization ability of the obtained domain-embeddings, which plays a vital role in improving the performance of the Logistic Regression classifier trained with relatively small amounts of labeled data.

\subsection{Post-processor}
The classification results from Detector can be used in two ways. First, we can assume that a domain name whose score is above a pre-set threshold is a DGA-generated domain name with a high probability. Thus, with this assumption, we can construct a feedback loop to update the blacklists and whitelists in the pre-processing session. Second, we can make use of the results of test dataset to conduct performance evaluation of the detector and make analysis on hard-cases, which is extremely meaningful for estimating the trends of current and in-coming DGAs and further improving the performance of DGA detection system.

\section{Datasets}

We collected DNS data for two consecutive weeks from the Tsinghua campus network using Passive DNS \cite{PassiveDNS} tools. A total of 162 million raw DNS query logs were obtained. The lengthy periods of data recording guarantee a representative dataset which contains different times of the day, different days of the week, and different working/non-working days. More information about the datasets is shown in Table 1.

There are some steps to be done before experiments. The critical points of our ideas are first filtering data with black/white lists and then manual labeling. Due to the huge amount of the collected raw data, we cannot afford to label them all. Thus we sample and label the first 15\% of the total 162 million queries and split this labeled dataset into two parts: 80\% as \emph{trainset-with-gt} (i.e. training set with ground truth) and the left 20\% as \emph{testset-with- gt} (i.e. test set with ground truth), which are employed to conduct the comparison experiments between our method and existing methods.

The reason why we choose the first 15\% of datasets for labeling is that during the time of collecting this part of data, we coincidentally found there are a large number of domains collected from query logs appearing on the malicious domain lists of some public blacklists such as DGArchive\cite{dgarchive}. Therefore, we started to collect data from that point of time and took these DNS logs containing DGA domains as the ground truth datasets to be labeled.

Meanwhile, the last 85\% unlabelled logs are used as the validation data for incremental word2vec algorithm because we intend to demonstrate that incremental word2vec could function well not only for the initial labeled datasets but also for newly added data. We randomly sample them at a scale of 1/10 in a consecutive way, then filter and manually label this new dataset. Again, the first 80\% and last 20\% of the dataset are put into \emph{trainset-with-gt} and \emph{testset- with-gt} respectively.

\begin{table}[htbp]
\caption{Details of TUNET DNS Datasets}
\begin{center}
\begin{tabular}{ll}
\toprule
\textbf{Properties}& \textbf{Descriptions} \\
\midrule
Duration of Data Collection & A total of consecutive 14 days \\
Generation Rate of DNS Queries & About 500 thousand / h \\
Peak Rate of DNS Queries & About 3 million / h \\
Occupied Space & About 7GB / day \\
Total Amount of DNS queries & About 162 million in total \\
Amount of Unique Domains Queried & About 1.9 million in total \\ 
\toprule
\end{tabular}
\label{tab1}
\end{center}
\end{table}

\begin{table}[htbp]
\caption{Analysis Results of Datasets with Ground-Truth}
\begin{center}
\begin{tabular}{|l|l|l|}
\toprule
\multicolumn{2}{|l|}{\textbf{Properties}}& \textbf{Values} \\
\midrule
All DNS  & Total Amount & 38,235,023 \\
\cline{2-3} 
Domains & Unique Amount  & 463,030 \\
\midrule

Benign DNS & Total Amount & 35,833,755 \\
\cline{2-3} 
Domains & Unique Amount  & 368,793 \\
\midrule

 & Total Amount & 2,401,268 \\
\cline{2-3} 
DGA DNS & Unique Amount  & 94,237 \\
\cline{2-3} 
Domains & Character-Based  & 84,538 (Unique) \\
\cline{2-3} 
 & Wordlist-based & 9,699 (Unique)\\
\toprule
\end{tabular}
\label{tab1}
\end{center}
\end{table}

\section{Experiments}

This section evaluates the performance of the proposed scheme over the real-world network. All operations were performed on a terminal server with Intel i7 8700K@3.70GHz CPU and 32GB RAM running Ubuntu Linux 16.04.

\subsection{Visualization}

In order to be intuitive, we use t-SNE \cite{tsne} to visualize the domain-embeddings generated from the incremental word2vec model. As is shown in Fig.2, DGA domains belonging to different families are labeled as class \textbf{1} to \textbf{11} and drawn in different colors, while benign domains are labeled as class \textbf{0} and densely clustered in the right of the picture. Among the classes, class \textbf{6} drawn in cyan refers to wordlist-based DGA domains, and the other classes except class 0 represent character-based DGA domains.

Fig.2 demonstrates that the clusters of malicious and benign domains can be divided neatly without difficulty with the help of incremental word2vec. It is suggested that those wordlist-based DGA domain names, which always mislead traditional methods, could be easily identified using incremental word2vec algorithm.

\begin{figure}[htbp]
\centerline{\includegraphics[width=3.3in]{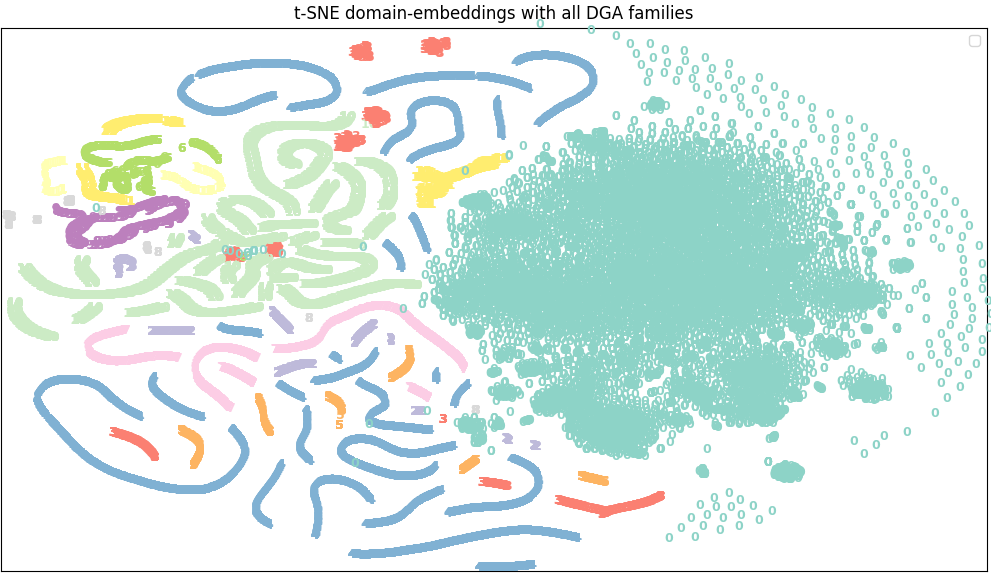}}
\caption{Visualization of Domain-Embeddings with all DGA Families Using t-SNE}
\label{fig}
\end{figure}

\begin{table}[htbp]
\caption{Results of Exp.1 to Exp.5 with All DGA Families}
\begin{center}
\begin{tabular}{c|cccccc}
\toprule
\textbf{Methods}& \textbf{Trainset}& \textbf{Testset}& \textbf{PRE}& \textbf{TPR}& \textbf{FPR}& \textbf{F1-score}  \\
\midrule
\emph{\textbf{FANCI}} & \emph{NXD} & \emph{NXD} &  \emph{0.791} & \emph{0.932} & \emph{0.117} & \emph{0.856} \\
%\midrule
\textbf{FANCI} & NXD & all &  0.001 & 0.176 & 0.603 & 0.001 \\
\textbf{FANCI} & all & all &  0.967 & 0.717 & 0.012 & 0.823 \\
\textbf{CNN} & all & all & 0.962 & 0.906 & 0.0004 & 0.934 \\
\textbf{LSTM} & all & all & 0.949 & 0.962 & 0.0003 & 0.947 \\
\textbf{IWM} & all & all & \textbf{0.989} & \textbf{0.980} & \textbf{0.0001} & \textbf{0.983} \\
\toprule
\end{tabular}
\label{tab1}
\end{center}
\end{table}

\begin{table}[htbp]
\caption{Results of Exp.1 to Exp.5 with only Wordlist-Based DGA}
\begin{center}
\begin{tabular}{c|cccccc}
\toprule
\textbf{Methods}& \textbf{Trainset}& \textbf{Testset}& \textbf{PRE}& \textbf{TPR}& \textbf{FPR}& \textbf{F1-score}  \\
\midrule
\emph{\textbf{FANCI}} & \emph{NXD} & \emph{NXD} &  \emph{0.014} & \emph{0.177} & \emph{0.118} & \emph{0.026} \\
%\midrule
\textbf{FANCI} & NXD & all &  0.001 & 0.002 & 0.987 & 0.001 \\
\textbf{FANCI} & all & all &  0.096 & 0.133 & 0.012 & 0.112 \\
\textbf{CNN} & all & all & 0.239 & 0.475 & 0.014 & 0.318 \\
\textbf{LSTM} & all & all & 0.345 & 0.489 & 0.010 & 0.403 \\
\textbf{IWM} & all & all & \textbf{0.996} & \textbf{1.000} & \textbf{0.000} & \textbf{0.995} \\
\toprule
\end{tabular}
\label{tab1}
\end{center}
\end{table}

\subsection{Comparison Experiment}

To further evaluate our system performance, we conduct comparison experiments between IWM (short for \emph{Incremental Word2Vec Model}), FANCI (representative of \emph{traditional machine learning methods}), CNN and LSTM (representative of \emph{character-based deep learning methods}).

It is notable that current popular DGA detection methods such as FANCI \cite{schuppen2018fanci} and D3N \cite{d3n} usually conduct experiments solely on NXDomains, which would miss many DGA domains. In fact, the number of DGA domains in NXDomains only accounts for less than 36\% of the total in all DNS data. Moreover, we find that FANCI method using Random Forests trained on NXDomains does not perform well on our testset containing only NXDomains, especially for wordlist-based malicious domains, which are almost undetectable in this case. Furthermore, if we evaluate the model with our testsets containing domains beyond the NXDomains, barely can it function normally.

Considering the situation above, we design the following sub-experiments. The results are published in Table 3 and Table 4.

\textbf{Exp.1.} RF (short for Random Forests) of FANCI, trained on NXDomains extracted from \emph{trainset-with-gt}, evaluated on NXDomains extracted from \emph{testset-with-gt}.

\textbf{Exp.2.} RF of FANCI, trained on NXDomains extracted from \emph{trainset-with-gt}, evaluated on \emph{testset-with-gt}.

\textbf{Exp.3.} RF of FANCI, trained on \emph{trainset-with-gt}, evaluated on \emph{testset-with-gt}.

\textbf{Exp.4.} CNN and LSTM, trained on \emph{trainset-with-gt}, evaluated on \emph{testset-with-gt}.

\textbf{Exp.5.} IWM, trained on \emph{trainset-with-gt}, evaluated on \emph{testset-with-gt}.

To be clear, we describe the training strategy for IWM here: As for labeled datasets, we extract the first half of the \emph{trainset-with-gt} as the initial train set and the last half as the new data continuously collected in the real world, named \emph{incremental-trainset}. And for unlabelled datasets used to train word embeddings, the same operations are conducted. For the sake of convenience, we divide the incremental train set (both labeled and unlabelled) into ten pieces. During the experiment, we first train the initial model on the initial train set and then conduct model updating operations for each newly added dataset.

The performance of models are evaluated with metrics as follows: \emph{Precision}, \emph{True positive rate} (also called \emph{Recall}), \emph{False positive rate} and \emph{F1-score}.

Through Exp.1,2 and 5, we can conclude that the performance of FANCI can hardly meet people’s expectations. And in order to validate that our method is superior to traditional machine learning and deep learning algorithms based solely on domain name strings with the same train set and test set, we conducted experiments 3, 4, and 5. We can see that IWM, whether tested with all DGA families or only wordlist-based DGA family, performs apparently better than FANCI, CNN, and LSTM.

What is more, it can be inferred that when confronted with wordlist- based DGA domain names, IWM trumps other detectors with the result of 100\% recall and 99.6\% precision, while the best of the others can hardly achieve half of the values.

\begin{figure}[htbp]
\centerline{\includegraphics[width=3.0in]{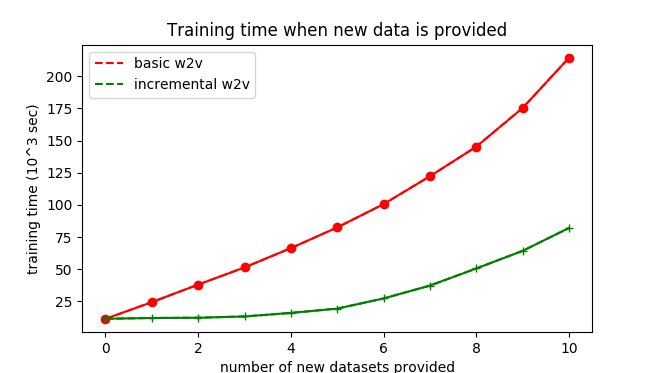}}
\caption{Training Time of Basic and Incremental Word2Vec Method When New Data is Provided)}
\label{fig}
\end{figure}

\begin{table}[H]
\caption{Comparison Results of Basic and Incremental Word2Vec Algorithm}
\begin{center}
\begin{tabular}{c|cccc|c}
\toprule
\textbf{Methods}& \textbf{PRE}& \textbf{TPR}& \textbf{FPR}& \textbf{F1-score} & \textbf{Training Time} \\
\midrule
\textbf{Basic} & 0.938 & 0.974 & 0.0003 & 0.943 & 214,213s\\
\textbf{Incre} & 0.949 & 0.962 & 0.0003 & 0.947 & 81,032s\\
\toprule
\end{tabular}
\label{tab1}
\end{center}
\end{table}

\subsection{Evaluation of Incremental Methodology}

This evaluation experiment is used to show that our polished version of the basic word2vec algorithm, i.e., incremental word2vec, could perform as well as or even better than its predecessor while gaining a tremendous acceleration in model updating. 
We construct two control groups: one is standard word2vec method, and the other is incremental word2vec method. The way training data is processed is the same as Exp.5 in section B, and both groups use \emph{testset-with-gt} as evaluation dataset. The training epoch number for both groups is 200 and the evaluation results are listed in Table 5 and Fig. 3.

\section{Conclusion and Future Work}

In this paper, we presented a novel system using incremental word2vec algorithm, which leverages inter-domain relationships to detect DGA domains effectively with scalable capability. Our system performs ex- cellently when confronted with various DGA families, even with wordlist-based DGAs, which are almost invincible for traditional detectors.

Moreover, to make model updating faster when new data is continuously provided, we explore an incremental training strategy. In our empirical experiments, we demonstrate that our incremental word2vec method could not only outperform other detectors but also gain a tremendous acceleration in model re-training.

Since the datasets for training and evaluation are collected continuously from the real-world networks, it is evident that our system is an online system which deals with tens of thousands of DNS query streams with high accuracy and efficiency.

The limitation of this paper is that the vocabulary of incremental word2vec model could become very large when unlimited data pours in, even though we already take measures such as merging common suffixes of domains to shorten the length of the vocabulary. It is difficult because we cannot merely limit the max length of the vocabulary due to the possible absence of some domain’s embedding that may lead the classifier to fail to find a suitable vector representation. Besides, labeled datasets of domains are hard to obtain. In future work, we will pursue solutions to these problems and get a better detection model.

\section*{Acknowledgment}

We thank Mingkai Tong, other classmates and teachers for their valuable help. Additionally, we thank Information Technology Center of Tsinghua University for authorizing the use of their data in our experiments. This work is supported by the National Science and Technology Major Project under Grant No.2017YFB0803004.

\bibliography{refer}{}
\bibliographystyle{IEEEtran}

\end{document}